# POSSIBLE OVERIONIZATION OF C II, N II, AND O II IONS IN THE ATMOSPHERES OF EARLY B- AND LATE O-TYPE STARS

L. S. Lyubimkov

*The reliability of C, N, and O abundances determined from lines of C II, N II, and O II in the main-sequence (MS) early B- and late O-type stars is examined. This analysis is based on a fact of primary importance for this problem: the C II, N II, and O II ions in the atmospheres of these stars are photoionized by radiation in the far UV. Observations show that the actual flux in this range for hot stars can be considerably higher than theoretical levels; thus, overionization of the C II, N II, and O II ions beyond the ionization calculated with the standard model atmospheres occurs in the actual atmospheres of these stars. Underestimates in the calculated ionization lead to reduced values of the C, N, and O abundances. This is confirmed by the observed dependence on the effective temperature $T_{eff}$: the C, N and O abundances tend to be lower with increasing $T_{eff}$. It is shown that overionization becomes significant for sufficiently hot stars, specifically when $T_{eff}>18500$ K for the C II lines and when $T_{eff}>26000$ K for the N II and O II lines. The systematic difference in the C, N and O abundances between these relatively hot stars and the cooler B-type MS stars is about 0.2 dex. The relatively cool B-type MS stars ($T_{eff}<18100$ K for C II and $T_{eff}<25000$ K for N II and O II) yield the undistorted C, N and O abundances, and their average values agree very well with modern estimates for the sun. This confirms that the metallicity of B- and O-type MS stars in the Sun's vicinity is the same as that of the sun. An explanation is given for the slight deficit of carbon found for the early B-type MS stars in previous work. It is noted that when an alternative method based on examining the ionization balance (e.g., He I-II, C II-III, Ne I-II, Si III-IV) is used to determine the parameters $T_{eff}$ and log g, the overionization problem essentially vanishes; however, this method leads to a systematic enhancement in the values of $T_{eff}$ and log g. The overionization problem can be solved by shifting to an*

---

Scientific-research Institute Crimean Astrophysical Observatory, Ukraine; e-mail: lyub@crao.crimea.ua

  

*improved model for the atmospheres of early B- and late O-type stars that takes their sphericity into account and, possibly, stellar winds and magnetic fields.*

Keywords: *stars: chemical composition*

## 1. Introduction

The stellar abundances of three light chemical elements (carbon, nitrogen, and oxygen) have been of great research interest for several decades. This is explained by the fact that, in terms of their abundance in stars, the elements C, N, and O (atomic numbers Z = 6, 7, and 8) are second only to hydrogen and helium. (Neon, Z = 10, also has an abundance close to that of C, N, and O.) Thus, an important parameter of stars, their metallicity, depends strongly on the abundances of these elements. In addition, these elements participate in the CNO cycle, which is the main source of energy in stars with masses $M > 2 M_\odot$ in the first and longest stage of their evolution, the main sequence (MS), during which hydrogen burns in the interior of a star. During the CNO cycle the amounts of C, N, and O inside a star vary greatly. After some time, because of mixing (owing, for example, to rapid rotation of a star in the MS stage) the C, N and O abundances can also vary in the surface layers. It follows from the theory of stellar evolution that after its completion, the MS phase in stars with $M > 2 M_\odot$ is rapidly followed by a phase of A-, F-, and G-supergiants, during which deep convective mixing takes place with additional variations in the surface abundances of C, N, and O. Thus, it is important to analyze the observed abundances of C, N, and O in stellar atmospheres in order to verify and improve the theory of stellar evolution.

Lines of carbon, nitrogen, and oxygen are observed in the spectra of stars of all spectral classes from O to M. In particular, lines of the C II, N II, and O II ions, which are usually employed to determine the C, N, and O abundances, can be seen in the spectra of early B- and late O-stars. In the 1990's an analysis of these lines for the early B-type MS stars in the sun's neighborhood showed that these stars have a small (~0.2-0.3 dex) deficit of C, N, and O relative to the sun [1-4]. We note that departures from local thermodynamic equilibrium (LTE) were taken into account in these calculations. This conclusion was unexpected from the standpoint of the theory of stellar evolution. In fact, calculations for the CNO cycle show that a deficit in C should have been accompanied by an excess of N, which was not observed.

The situation changed in the 2000's mainly because of revisions with a reduction in the solar abundances of C, N, and O. This happened because an improved model for the sun's atmosphere was used: a nonstationary hydrodynamic 3D model which reproduces the solar granulation. The abundances of all elements, including C, N, and O, were revised [5]. In addition the accuracy of the determination of the C, N and O abundances was increased, also for the B-type MS stars. The most recent work on these stars [6,7] shows that the average abundances of nitrogen and oxygen in B-type MS stars are now in good agreement with modern estimates for the sun. However, there is still a small deficiency for C. For example, an average carbon abundance of $\log\varepsilon(C) = 8.33 \pm 0.04$ has been obtained for a sample of 29 B-type MS stars [6], and a close value of $\log\varepsilon(C) = 8.31 \pm 0.13$ for another sample of 22 B-type MS stars [7]. These are markedly lower than current estimates for the sun: $\log\varepsilon(C) = 8.43 \pm 0.05$ [5] and [8], especially given the very small errors in the determinations of $\log\varepsilon(C)$ in Refs. 6, 5, and 8. It should be noted that the B-type MS stars examined in Refs. 6 and 7 have quite slow rotation, and the initial amounts of C, N, and O appear to be retained in their atmospheres. We note that the amount of carbon and other elements is given here on a normal logarithmic scale, with $\log\varepsilon(H) = 12.00$ taken for hydrogen.

It is important to note that in this paper we are discussing the observed abundances of C, N, and O for rather young B-type MS stars in the sun's neighborhood. In fact, their age does not generally exceed 100 million years, or considerably less than the sun's age of 4.5 billion years. The distances to these stars [7] do not exceed 600 pc. Is it possible that the initial abundances of metals in such young and nearby stars could differ from the metal content of the sun? (We recall that in astrophysics all elements heavier than hydrogen and helium are referred to as metallic; usually the "metallicity" of a star, i.e., its metal content, is defined relative to the sun.)

In model calculations of the chemical evolution of the galaxy it has traditionally been assumed that the sun has remained in the same region of the galaxy in which it was formed 4.5 billion years ago. However, some recent calculations provide an alternative view, according to which the sun was substantially closer to the center of the galaxy when it was formed than now. In that case its chemical composition may differ from that of the young stars in its current environment [6]. It should be noted that, on one hand, calculations of the migration of stars (including the sun) in the galactic disk are quite difficult and are still quite uncertain. On the other hand, the amounts of many metals in nearby young stars agree well with the solar abundances; this does not support the alternative hypothesis. For example, agreement of this sort has been found between nearby B-type MS stars and the sun for the following elements: N and O [6,7], Mg [9], and Mg, Si, and Fe [10]. In the case of class A, F, and G supergiants, which are the direct descendents of B-type MS stars, agreement with the sun has been obtained for such indicators of metallicity as Fe [11] and Cr and Ti [12].

Thus, the available empirical data on the chemical composition of young nearby stars confirms that the metallicity of these stars is the same as that of the sun. Only a small, but stable deficit of carbon in the atmospheres of B-type MS stars appears to conflict with the overall pattern. It seems possible that carbon, like the other metals, may have a normal (solar) abundance, but the observed deficit in it is the result of imperfections in the methods being used. Here we attempt to justify this proposition. It will be shown that a number of facts support the possible overionization of C II, N II, and O II ions in the atmospheres of early B-type MS stars, as well as of hotter O-stars. Here we keep in mind that the degree of ionization in actual atmospheres is considerably higher than in calculations based on standard atmospheric models. In other words, overionization of C II, N II, and O II has not been taken into account in the calculations and this may lead to reduced values for the C, N and O abundances.

## 2. The adequacy of the models for stellar atmospheres

The calculations of the C II, N II, and O II lines that form the basis of determinations of the C, N and O abundances in early B- and late O-stars have been done in the last two decades without assuming LTE. Standard atmospheric models similar to those of Kurucz [13] are used in these non-LTE calculations of the C II, N II, and O II lines; that is, these are plane-parallel LTE models in hydrostatic equilibrium. To what extent do they correspond to the observed characteristics of the O- and B-stars?

The first indication that the standard model atmospheres are not entirely adequate for the actual atmospheres of O- and early B-stars can be regarded as the discovery of x-ray emission from normal O- and B-stars. X-ray fluxes from these stars have been discovered and studied with the aid of the EINSTEIN, ROSAT, and CHANDRA satellites [14-16]. The x-ray luminosity $L_x$ of these stars is low: its ratio to the bolometric luminosity $L_{bol}$ is typically $L_x/L_{bol} \sim 10^{-7}$. Nevertheless, the presence of the x-rays cannot be explained by the standard models of stellar atmospheres. To explain them it was necessary to invoke either a hot corona with a temperature $T \geq 10^6$, or a strong stellar wind. Thus, the standard models for the atmospheres of O- and B-stars are unable to predict the observed x-ray fluxes. It has come to be assumed that the very process that leads to overheating of the outer layers of the atmospheres of hot stars and to the appearance of the x-rays (e.g., shock waves in a stellar wind [16]) can greatly increase the hard radiation in the far ultraviolet (UV); it will be shown below that the latter plays an important role in the calculations for the C II, N II, and O II lines.

The intensity of the C II, N II, and O II lines in the spectra of hot O- and B-stars depends on the concentration of the C II, N II, and O II ions in their atmospheres. A correct non-LTE analysis of these lines involves, in particular, a correct treatment of the ionization processes for these ions. Two ionization processes have to be taken into account in solving the non-LTE problem: collisional ionization and ionization under the influence of radiation, i.e., photoionization. It turns out that photoionization by far UV radiation is of primary importance for the C II, N II, and O II lines in the case of the O- and early B-stars.

Table 1 lists the ionization potentials $E_{ion}$ of the C II, N II, and O II ions and, for comparison, those of the H I and He I atoms. The ionization limits, i.e., the wavelengths $\lambda_0$ (Å) corresponding to ionization from the ground state are also given here. It is clear from Table 1 that photoionization of the C II, N II, and O II ions is controlled by radiation in the Lyman continuum of H I, which corresponds to wavelengths $\lambda < 912$ Å, and, additionally, for the N II and O II ions by He I continuum emission, i.e., emission with $\lambda < 504$ Å. Until recently, calculations of the photoionization of C II, N II, and O II ions have used theoretical UV fluxes derived from standard models of stellar atmospheres. The question arises of the extent to which these fluxes correspond to the observed far UV emission for hot O- and B-stars.

TABLE 1. Ionization Potentials and the Corresponding Wavelength Limits for Five Light Elements

| Atom or ion | $E_{ion}$, eV | $\lambda_0$, Å |
|---|---|---|
| H I | 13.595 | 912.0 |
| He I | 24.587 | 504.3 |
| C II | 24.383 | 508.5 |
| N II | 29.601 | 418.9 |
| O II | 35.117 | 353.1 |

Unfortunately, observations of stars in the region of the Lyman continuum ($\lambda < 912$ Å) are very difficult because of strong absorption by interstellar hydrogen. Nevertheless, the EUVE satellite, launched in 1992, has yielded the distribution of the flux for $\lambda < 912$ Å for two early B-stars, $\varepsilon$ CMa (B2 II) and $\beta$ CMa (B1 II-III) [17]. These stars happened to lie in a fairly transparent region of space, where the number of interstellar hydrogen atom absorbers is comparatively low, so it was easier to detect the UV flux. For one of the stars, $\varepsilon$ CMa, it has also been possible to measure the UV continuum flux of He I ($\lambda < 504$ Å) [18]. A comparison with the theoretical distribution of the UV flux calculated with Kurucz models [13] showed that the observed flux is greater than the theoretical flux by two orders of magnitude for $\lambda < 504$ Å and by several times for $\lambda < 912$ Å (see Fig. 12 of Ref. 19 and Fig. 2 of Ref. 20). This implied that standard models of stellar atmospheres similar to the models of Ref. 13 cannot explain either the x-ray emission in O- and early B-stars, or the observed far UV flux.

Aufdenberg, et al. [20,21], show that spherical non-LTE models for the atmospheres of B-stars provide a much better explanation of the observed distribution of the UV flux for the star $\varepsilon$ CMa than the standard plane-parallel models. Here the deviations from LTE play a secondary role, with the major contribution from the transition to a spherical, extended atmosphere from a plane geometry. Later, more satisfactory agreement was obtained [22] with the observations on $\varepsilon$ CMa; however, although the level of the calculated UV flux agreed better on the average with the observations, the wavelength distributions of the flux in the calculations and the observations were substantially different. In addition, the observed intensities of the spectrum lines in the UV region under consideration were too strong compared to the calculations [22].

In other words, today's models of the atmospheres of early B-stars still cannot provide an adequate description of the observed far UV flux. It is clear from the above discussion that the use of spherical models of the atmosphere for O- and early B-stars would seem to be preferable for examining the photoionization of C II, N II, and O II ions. However, all the work on determining the C, N and O abundances in B- and O-stars has, as before, been based on applying standard plane-parallel model atmospheres. These models are inadequate for modelling the far UV flux and may lead to reduced ionization of C II, N II, and O II and, therefore, to excessively low values for the C, N and O abundances.

## 3. Observed abundances of C, N, and O: the dependence on $T_{eff}$.

What are the consequences of underestimating the effect of photoionization? As an example, consider Ref. 23, where this effect is examined for the Fe I lines in the spectra of F-supergiants. It was shown that the actual photoionization of Fe I atoms by UV radiation (here $\lambda_0 = 1575$ Å) is considerably higher than implied by the standard LTE models of stellar atmospheres. Including this overionization of Fe I in non-LTE calculations leads to an increase in the Fe abundance and brings the latter into agreement with the abundance determined using Fe II ion lines. A very similar situation arises for the hotter stars in classes O and B on analyzing lines of C II, N II, and O II. As noted above, far UV observations of several early B-stars suggest that in non-LTE calculations of the C II, N II, and O II lines for these stars the assumed UV flux is low, so that the derived abundances of C, N, and O are also reduced. Since the UV flux is higher for higher effective temperatures $T_{eff}$, we may expect that an underestimate of the photoionization of the C II, N II, and O II ions will show up more noticeably for stars with higher $T_{eff}$; that is, there is a dependence on $T_{eff}$.

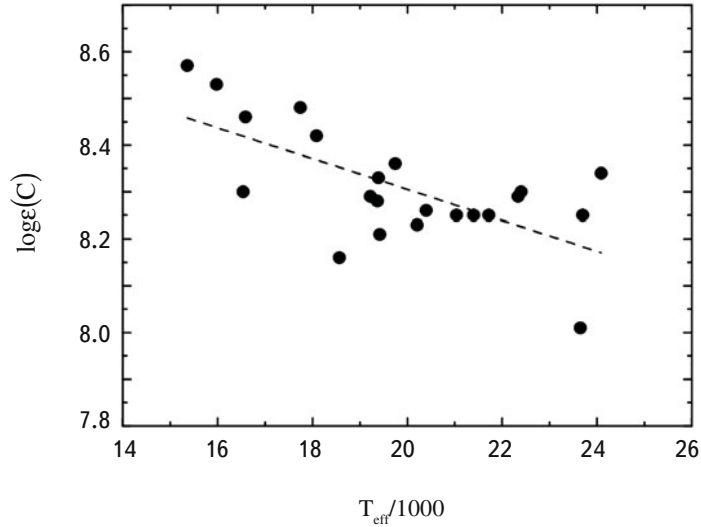

Fig. 1. Carbon abundance as a function of $T_{eff}$ for 22 B-stars [7]. The dashed line is a least squares fit.

We now examine observational data from several papers that confirm this assumption. As noted above, of the group of elements C, N, and O, a deficit of carbon is of greatest interest. Figure 1 shows the dependence of the abundance of carbon, $\log\varepsilon(C)$, on $T_{eff}$ for 22 B-stars based on data from Ref. 7. This figure shows that there is a trend in $\log\varepsilon(C)$ with $T_{eff}$, specifically, the relatively cool stars with effective temperatures $T_{eff} < 18100$ K have a higher abundance of C on the average than the stars with $T_{eff} > 18500$ K. The abundance of C for stars with $T_{eff} = 15300\text{-}18100$ K is essentially the same as the solar abundance, while it is systematically lower for stars with $T_{eff} = 18500\text{-}24100$ K. This can be seen in Table 2, where the average abundances of C for the relatively cool and hot B-stars [7] are compared. For the stars with $T_{eff} < 18100$ K the average abundance $\log\varepsilon(C) = 8.46 \pm 0.09$ is in outstanding agreement with modern estimates for the sun: $\log\varepsilon(C) = 8.43 \pm 0.05$ [5] and $\log\varepsilon(C) = 8.50 \pm 0.06$ [8]. On the other hand, the hotter stars with $T_{eff} > 18500$ K have a systematic deficit in C compared to the sun that averages about 0.2 dex.

TABLE 2. Average Abundance of Carbon for Relatively Cool and Hot B-stars [7]

| Objects | $\log\varepsilon(C)$ |
|---|---|
| 6 B-stars with $T_{eff} < 18100$ K | 8.46 ± 0.09 |
| 16 B-stars with $T_{eff} > 18500$ K | 8.25 ± 0.08 |

Stars with only a relatively narrow range of effective temperatures, $T_{eff}$ = 15300-24100 K, were examined in Ref. 7. As Fig. 1 shows, this turned out to be enough to see a trend in the abundance of C with $T_{eff}$. Such a limited range of $T_{eff}$ would be unsuitable for plotting the analogous dependences on $T_{eff}$ for oxygen or nitrogen, since photoionization of the O II and N II ions involves harder UV radiation (Table 1) and, therefore, requires higher temperatures for detection of a possible trend. The data of Daflon, et al. [24], are better suited for this sort of analysis, since they cover a range of $T_{eff}$ from 19000 to 34000 K, i.e., they include late O-stars, and not just early B-stars. It should be noted that the 15 stars studied by them [24] have low rotation velocities and, in the overwhelming majority of cases, are far from completing the MS state; hence, we may expect that, as in Ref. 7, their atmospheric abundances of C, N, and O will not have changed during their evolution on the MS.

Unfortunately, the carbon abundance was found for only 6 stars in Ref. 24, so it is not possible to construct a reliable dependence of $\log\varepsilon(C)$ on $T_{eff}$ here. The most complete data in that paper [24] are for oxygen (15 stars) and these are shown in Fig. 2. As in the case of carbon in Fig. 1, here there is an obvious anticorrelation between $\log\varepsilon(O)$ and $T_{eff}$. The dashed line is a least squares fit and indicates that the observed abundance of O decreases by about 0.4 dex when $T_{eff}$ is increased from 19000 to 34000 K. The stars included in Fig. 2 can be divided into two groups with relatively high and low abundances of O; the watershed between them lies at $T_{eff}$ = 25000-26000 K. According to Table 3, stars with $T_{eff}$ < 25000 K have an abundance in good agreement with that of the sun; the average for these stars, $\log\varepsilon(O) = 8.75 \pm 0.14$ is in outstanding agreement with current estimates for the sun, $\log\varepsilon(O) = 8.69 \pm 0.05$ [5] and $\log\varepsilon(O) = 8.76 \pm 0.07$ [25]. On the other hand, the hotter stars with $T_{eff}$ > 26000 have a systematic deficit in O relative to the sun that averages about 0.2 dex.

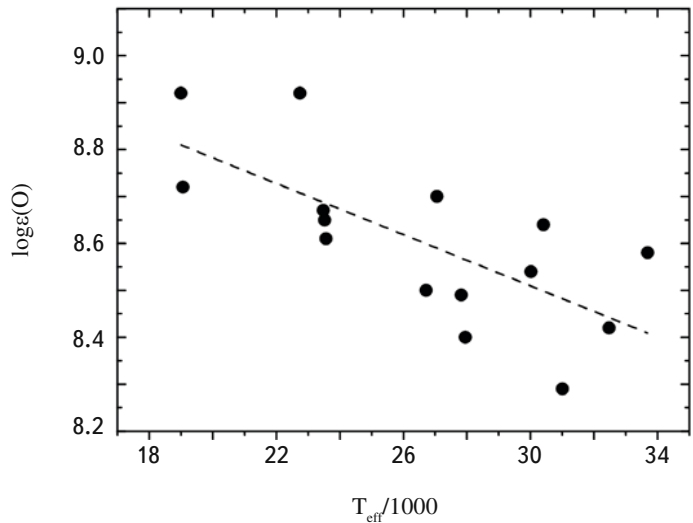

Fig. 2. Oxygen abundance as a function of $T_{eff}$ for 15 B- and O-stars [24]. The dashed line is a least squares fit.

TABLE 3. Average Abundance of Oxygen for Relatively Cool and Hot B-stars [24]

| Objects | $\log\varepsilon(O)$ |
|---|---|
| 6 B-stars with $T_{eff}$ < 25000 K | 8.75 ± 0.14 |
| 9 B-stars with $T_{eff}$ > 26000 K | 8.51 ± 0.13 |

The abundance of nitrogen has been determined for 13 stars [24], i.e., for a smaller number of stars than the abundance determination for oxygen. In addition, the accuracy of the data is somewhat lower here. Nevertheless, as Fig. 3 shows in this case there is also a trend with $T_{eff}$. The least squares fit line shows that the drop in the observed abundance of N as $T_{eff}$ increases from 19000 to 34000 K is 0.22 dex. According to Table 4, the average abundance $\log\varepsilon(N) = 7.80 \pm 0.16$ for the stars with $T_{eff}$ < 25000 K is in very good agreement with current values for the sun, $\log \varepsilon(N) = 7.83 \pm 0.05$ [5] and $\log \varepsilon(N) = 7.86 \pm 0.12$ [26]. On the other hand, for the stars with $T_{eff} > 26000$ K there is a deficit of about 0.2 dex of N relative to the sun.

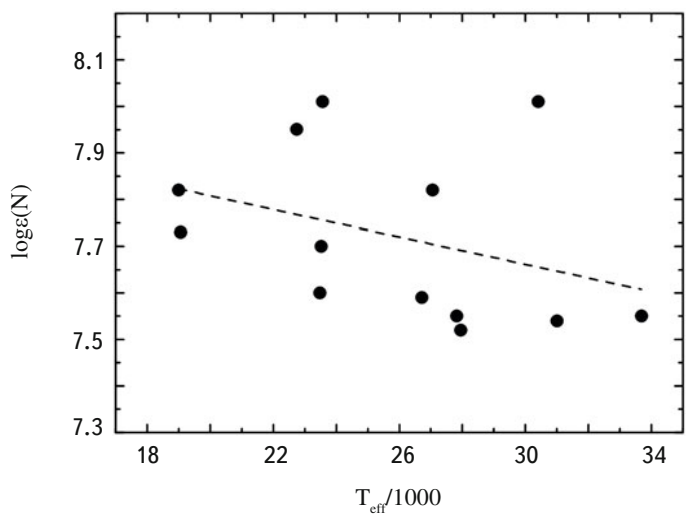

Fig. 3. Nitrogen abundance as a function of $T_{eff}$ for 13 B- and O-stars [24]. The dashed line is a least squares fit.

TABLE 4. Average Abundance of Nitrogen for Relatively Cool and Hot B-stars [24]

| Objects | $\log\varepsilon(N)$ |
|---|---|
| 6 B-stars with $T_{eff}$ < 25000 K | 7.80 ± 0.16 |
| 7 B-stars with $T_{eff}$ > 26000 K | 7.65 ± 0.19 |

An analysis of Refs. 7 and 24 shows, therefore, that the C, N and O abundances for relatively cool B-type MS stars are the same as for the sun; this applies to stars with $T_{eff} < 18100$ K in the case of C and $T_{eff} < 25000$ K for N and O. The hotter stars ($T_{eff} > 18500$ K for C and $T_{eff} > 26000$ K for N and O) have a systematic deficit in C, N, and O averaging 0.2 dex relative to the sun. (The deficit was roughly the same for all three elements.) This deficit may be the result of underestimating the degree of ionization in the non-LTE calculations for the C II, N II, and O II lines in the case of the comparatively hot stars because of a lower ionizing UV flux. In other words, overionization of the C II, N II, and O II ions may not have been taken into account in the calculations.

It follows from the above discussion that overionization becomes significant in the case of C II when $T_{eff} > 18500$ K and in the case of N II and O II when $T_{eff} > 26000$ K. It is interesting to note that, according to non-LTE calculations for C II lines, the maximum equivalent widths of these lines are reached when $T_{eff} = 20000\text{-}22000$ K; for N II lines the maximum is reached when $T_{eff} = 24000\text{-}26000$ K and for O II, when $T_{eff} = 26000\text{-}28000$ K [7]. Thus, overionization becomes noticeable near values of $T_{eff}$ such that the corresponding lines are strongest. As $T_{eff}$ increases further these lines become weaker, possibly because of the decreasing C II, N II, and O II concentrations in stellar atmospheres.

It should be emphasized that in studies of the chemical composition of stars the accuracies of two fundamental parameters are crucial: the effective temperature $T_{eff}$ and the acceleration of gravity $\log g$. As noted in Ref. 7, various authors have different $T_{eff}$ scales for early B-stars; the differences are especially noticeable for the hottest B-stars with $T_{eff} \geq 25000$ K. For this reason, relatively cool stars with $T_{eff} \leq 24100$ K, for which $T_{eff}$ is determined more accurately, are examined in Ref. 7. (Nevertheless, this was sufficient to detect a trend for carbon with $T_{eff}$; see Fig. 1.)

For further discussion it is necessary to recall briefly the method for determining the parameters $T_{eff}$ and $\log g$ in the earlier papers [7,24] that have been used to construct Figs. 1-3. In particular, in Ref. 7 $T_{eff}$ was determined using two photometric indices that are free of interstellar absorption, specifically the index $Q$ in the *UBV* system and the index $[c_1]$ in the *uvby* system. An improved method based on stellar parallaxes was used [7] for determining $\log g$. (This method is described in Refs. 27 and 11.) The observed parallaxes were taken from a new reduction of HIPPARCOS data [28]. This led to unprecedented accuracy in determining $\log g$ for these fairly nearby stars [11]. As for Ref. 24, data from which has been used for the plots of Figs. 2 and 3, $T_{eff}$ was determined using the index $Q$ of the *UBV* system, along with several photometric indices in the *uvby* system, and $\log g$ was estimated using the profile of the Balmer H$\gamma$ line.

## 4. Artificial enhancement of the ionization because of the parameters $T_{eff}$ and $\log g$

The problem of a dependence of the C, N and O abundances in early B-type MS stars on their effective temperature $T_{eff}$ has been encountered previously. For example, twenty years ago Gies and Lambert [2] found a trend in the C, N and O abundances with increasing $T_{eff}$ (see Fig. 11 in Ref. 2). It is important to note they used a method for estimating $T_{eff}$ and $\log g$ that is similar to that used in Refs. 7 and 24 cited above (photometric indices and Balmer lines were used). In order to eliminate this trend, these authors introduced a correction of 3.4% in their value of $T_{eff}$. An enhancement in $T_{eff}$ of this sort evidently led to an increased UV flux in model atmospheres of the stars being studied and, therefore, to increased photoionization of the C II, N II, and O II ions that is especially noticeable for the hottest B-stars. This, in turn, caused a drop in the intensity of the calculated C II, N II, and O II lines and an increase in the resulting abundances of C, N, and O which was more noticeable for higher $T_{eff}$. With this artificial enhancement in $T_{eff}$ it was possible to eliminate the trend in the C, N and O abundances with increasing $T_{eff}$; that is, any need for overionization of the C II, N II, and O II ions was removed implicitly.

An entirely different method of determining $T_{eff}$ and log $g$ has been used in some recent papers. An example is Ref. 6, in which the C, N and O abundances and four other elements were determined for 20 B-stars with Teff ranging from 15000 to 34000 K. All of these stars are in the MS stage and have low observed rotation velocities ($v\sin i \leq 30$ km/s). It would seem that this wide range of $T_{eff}$ would be appropriate for a search for dependences analogous to those shown in Figs. 1-3. However, no trend in the C, N and O abundances was found [6].

Reference 6 differs fundamentally from Refs. 7 and 24 in the method for determining $T_{eff}$ and log $g$, which is based here exclusively on an examination of the ionization equilibrium (photometric indices were not considered). This method for the ionization balance assumes that values of $T_{eff}$ and log $g$ have been chosen for each star such that the abundances of a given element found for two neighboring ionization states must be the same. Lines of the following elements in two ionization states were used: He I-II (for $T_{eff}$ > 26000 K), C II-III, O I-II (for $T_{eff}$ < 24000 K), Ne I-II (for $T_{eff}$ > 22000 K), Si III-IV, and Fe II-III (for $T_{eff}$ < 24000 K). In particular, the pairs Ne I-II and Si III-IV were examined along with He I-II and C II-III for relatively hot stars. As in the case of He I and C II (Table 1), photoionization of Ne I and Si III is driven by far UV radiation; i.e., here, also, overionization is possible. In fact, the ionization potential of Ne I is 21.56 eV and that of Si III is 33.49 eV; these values correspond to ionization limits of $\lambda_0$ = 575 and 370 Å. This means that compensation for the overionization effect was implicit in the method used for determining $T_{eff}$ and log $g$ in Ref. 6.

In particular, for 18 of the 20 stars studied in Ref. 6 the abundance of carbon was determined using C II and C III lines; only for the two coolest stars (with $T_{eff}$ = 15800 and 17500 K) were C II lines alone used. The same pair of ions, C II and C III, were used for determining the parameters $T_{eff}$ and log $g$ for these stars. These parameters were chosen so that the average abundance of C found from the C II lines was the same as that found from the C III lines. Because of the high sensitivity of the ionization of C II to $T_{eff}$, a slight increase in $T_{eff}$ and the associated variation in log $g$ meant that the overionization effect was undetectable. In fact, if the abundance of C is plotted as a function of $T_{eff}$ for all 20 stars from Ref. 6, no trend of the sort seen in Fig. 1 can be found.

The abundance of nitrogen in Ref. 6 has a large spread, about 0.5 dex (it was determined [6] only using lines of N II). It is possible that the abundance of N in the atmospheres of some of the stars examined there [6] have increased during their evolution on the MS. With such a large spread, no visible trend in the abundance of N with $T_{eff}$ could be found.

The abundance of oxygen in Ref. 6, on the other hand, does have a slight trend with $T_{eff}$. (For the stars with $T_{eff}$ > 26000 K the abundance of O was found only using lines of O II, while for the cooler stars both O I and O II lines were used.) If a least squares linear fit is calculated as in Fig. 2, it turns out that the abundance of O decreases by 0.1 dex when $T_{eff}$ is raised from 15000 to 34000 K. This is considerably smaller than in Fig. 2; that is, we can assume that, in the case of O II, as opposed to C II, overionization was not fully compensated in Ref. 6. This can be explained fully by the fact that the O II ions are ionized by much harder UV radiation than the C II ions (see Table 1), so that the choice of the parameters $T_{eff}$ and log $g$ for the pair C II-C III may not give as good agreement for the pair O II-O III. In other words, complete elimination of the overionization effect for O II-O III would require a higher value of $T_{eff}$ than for C II-C III.

Therefore, the results of Ref. 6, in particular, the absence of a trend with $T_{eff}$ for C and the small "residual" trend for O, can be explained by the fact that the method used there [6] for determining $T_{eff}$ and log $g$ made it possible to avoid the overionization effect fully (in the case of C) or partially (in the case of O).

A similar method for determining $T_{eff}$ and log $g$, based on the ionization balance of Si II-III-IV has also been used by Morel, et al. [29], who examined 8 stars within a fairly narrow region of $T_{eff}$ = 22500-27500 K. Thus, as in Ref. 6, the method used by them [29] for estimating $T_{eff}$ and log $g$ obviously facilitated the elimination of a trend in the C, N and O abundances. In addition, here [29] both the number of stars and the range of $T_{eff}$ were clearly insufficient for a search for a trend with $T_{eff}$. The same method for determining $T_{eff}$ and log $g$ has been used in a study [30] of 18 stars in two clusters (NGC 3293 and NGC 4755) with $T_{eff}$ in the range from 16000 to 23500 K. A comparison with Figs. 1-3 shows that this range of $T_{eff}$ was too narrow; in addition, the method for estimating $T_{eff}$ and log $g$, as in Refs. 6 and 30, facilitated the elimination of any possible trend.

It is, therefore, possible that the ionization balance technique used to determine $T_{eff}$ and log $g$ in Refs. 6, 29, and 30 leads to a systematic enhancement in $T_{eff}$ and a corresponding variation in log $g$ compared to the traditional method used previously. In order to verify this hypothesis, we can compare the values of $T_{eff}$ and log $g$ from Ref. 6 with a determination of $T_{eff}$ and log $g$ by the method of Napiwotzki [31] based on using photometric indices in the $uvby$-β system. Here the index β characterizes absorption in the Balmer Hβ line; therefore, this method [31] can be regarded as one of the traditional methods for determining $T_{eff}$ and log $g$ in B-stars. It should be noted that in Ref. 7 very good agreement is found between the values of $T_{eff}$ obtained for B-stars with $T_{eff}$ ranging from 15300 to 24100 K based on the indices $Q$, $[c_1]$ and parallaxes, on one hand, and by the method of Ref. 31, on the other.

Calculations using the program of Ref. 31 for the 20 stars examined in Ref. 6 have been done by Rostopchin (personal communication). A comparison of the results with the data of Ref. 6 showed that the differences $\Delta T_{eff}$ and $\Delta \log g$ between the values of $T_{eff}$ and log $g$ found by the two methods reveal a systematic difference between hot B-stars with $T_{eff} \geq 24000$ K and cooler B-stars with $T_{eff} < 24000$ K; that is, a trend in the values of $\Delta T_{eff}$ and $\Delta \log g$ can be seen with increasing $T_{eff}$. It is interesting that the ionization balance technique used in Ref. 6 leads to a systematic enhancement in both $T_{eff}$ and $log\ g$ for the hot B-stars with $T_{eff} \geq 24000$ K. The value of $T_{eff}$ for these stars can be enhanced in Ref. 6 by up to 2300 K and the value of log $g$, by up to 0.5 dex. The consequence of the enhancement in both quantities, $T_{eff}$ and log $g$, for hot B-stars should be an enhancement in the ionization in the calculations for C II, N II, and O II and, ultimately, an increase in the C, N and O abundances that are determined.

Thus, there is an obvious contradiction. On one hand, if the traditional method based on photometry and the Balmer lines for determining $T_{eff}$ and log $g$ for early B- and late O-stars is used, then non-LTE calculations of the C II, N II, and O II lines will yield abundances of C, N, and O which manifest a trend with increasing $T_{eff}$. Overionization must be introduced in order to eliminate this trend. On the other hand, if the method for determining $T_{eff}$ and log $g$ is based exclusively on an examination of the ionization balance, then this trend vanishes and overionization is not necessary; however, here the values of $T_{eff}$ and *log g* for the hot stars are systematically higher than in the first case. Evidently, the reason for this contradiction is the use of standard model atmospheres, which are not fully adequate for describing actual stellar atmospheres.

## 5. Discussion

The trend in Figs. 1-3 becomes noticeable only for fairly hot stars. As for the relatively cool stars, their abundances logε have a spread that can be explained by errors in the determination and by small real variations in logε from star to star. Specifically, this refers to B-stars with effective temperatures $T_{eff}$< 18000 K in the case of carbon and $T_{eff}$< 25000 K in the case of nitrogen and oxygen. It can be assumed that the overionization effect for these stars is small, so, on one hand, they yield undistorted values of the C, N and O abundances. On the other hand, since we are dealing with stars that are far from completing the MS stage, these abundances of C, N, and O can be regarded as the initial values for these kinds of stars. Table 5 lists the average values of $\log\varepsilon(C)$, $\log\varepsilon(N)$, and $\log\varepsilon(O)$ for the relatively cool B-stars with the above values of $T_{eff}$ [7,24]. (Here partial use is made of data from Tables 2-4.) Two conclusions follow from Table 5: (1) in the case of N and O, for which both papers provide data, the agreement between Refs. 7 and 24 is good; (2) the results on the C, N and O abundances for these stars are in very good agreement with modern estimates for the sun (last column of Table 5). This confirms that the initial abundances of C, N, and O in B-type MS stars at the start of their evolution are the same, on the average, as the values for the sun.

It should be emphasized that, according to Table 5, there is no deficit in the initial abundance of carbon relative to the sun. Recall that in earlier papers a deficit of C was found for unevolved B-type MS stars, but this had no convincing explanation. It is shown above that when $T_{eff}$> 18500 K the calculations give a reduced ionization of C II, which leads to a lower abundance $\log\varepsilon(C)$. Thus, a reduced abundance of carbon was obtained over essentially the entire range of $T_{eff}$ = 18000-30000 K covered by the early B-type MS stars. The situation is somewhat more favorable for N and O, since overionization becomes noticeable for them only when $T_{eff}$> 26000 K.

TABLE 5. Average Initial Abundances of C, N, and O in B-type MS Stars in the Sun's Neighborhood [7, 24]

| Element | Number of stars | logε | Sun |
|---|---|---|---|
| C | 6 | 8.46 ± 0.09 [7] | 8.43 ± 0.05 [5] |
|   |   |   | 8.50 ± 0.06 [8] |
| N | 6 | 7.80 ± 0.16 [24] | 7.83 ± 0.05 [5] |
|   | 20 | 7.78 ± 0.09 [7] | 7.86 ± 0.12 [26] |
| O | 6 | 8.75 ± 0.14 [24] | 8.69 ± 0.05 [5] |
|   | 20 | 8.72 ± 0.12 [7] | 8.76 ± 0.07 [25] |

A preliminary estimate of the overionization can be made. As noted above, in order to eliminate the systematic difference in the C, N and O abundances between comparatively cool and hotter B-stars (see Tables 2-4), it is necessary to raise the C, N and O abundances for the hot stars by roughly 0.2 dex or a factor of 1.6. To do this, it is enough to reduce the populations of all levels of the C II, N II, and O II ions by a factor of 1.6 in the non-LTE calculations; then this difference between the hot and cool B-stars vanishes. The level populations can be reduced by increasing the degree of ionization of C II, N II, and O II, which, in turn, can be attained either by artificially increasing $T_{eff}$ and log $g$ (see above) or by increasing the intensity of the far UV ionizing radiation. In order to increase the theoretical UV flux, it is necessary to go to improved models for the atmospheres of B- and O-stars.

This estimate of the amount of overionization (a factor of ~1.6) is fairly crude. It should be kept in mind that the visible C II, N II, and O II lines of concern here are formed from quite high levels with excitation energies of 20 eV or more. It is possible that the overionization of lower levels and, especially, the ground states of these ions will be considerably higher. Clarification of this point requires detailed non-LTE calculations based on the improved model atmospheres that take into account the actual conditions in the atmospheres of B- and O-stars.

The improvements in the models will involve, first of all, the use of spherical instead of plane-parallel geometries. As noted above, spherical models for the atmospheres give a better description of the observed far UV radiative flux for early B-stars. The calculated UV flux in these models is higher because of a higher temperature in the region where the Lyman continuum is formed [20,21] and this should lead to increased ionization of the C II, N II, and O II ions. At present, there are several computer codes that can be used to calculate spherical model atmospheres of hot stars [32]. Unfortunately, as before, models with plane geometries are used to determine the C, N and O abundances based on lines of C II, N II, and O II. Another improvement, at least for hot giants and supergiants, would be to include stellar wind in the calculations, as in the code FASTWIND [33]. We note that stellar winds (more precisely, shock waves in them [32]) are now regarded as a necessary condition for the production of x-radiation in hot stars. Finally, magnetic fields on the order of hundreds of Gauss or even around 1500-1700 G have been detected in some O- and early B-stars [34]. Magnetic fields may also have a certain role in the structure of the atmosphere of these stars. Thus, once the effects listed here (sphericity, stellar wind, and magnetic fields) were included in calculations for model atmospheres, they could aid in solving the overionization problem.

In discussing the role of overionization in determining the C, N and O abundances in hot stars, we should consider, however briefly, yet another light element which is listed in Table 1. That is helium, the abundance of which is found using lines of He I. Since photoionization of He I atoms, as that of C II, N II, and O II ions, is driven by far UV radiation ($\lambda < 504$Å, see Table 1), it may be necessary to introduce overionization into the relevant calculations. In order to test this assumption, we can attempt to find a trend in the abundance of helium with $T_{eff}$ similar to that found above for the C, N and O abundances (Figs. 1-3). For this purpose, we use some non-LTE estimates of the helium abundance He/H for 102 B-type MS stars [35]. He/H is the "helium/hydrogen" ratio of the number of atoms. The effective temperatures $T_{eff}$ and other parameters for these stars were determined in Ref. 36, with $T_{eff}$ and log $g$ found by the traditional method (using photometry and Balmer lines). All these stars are in the MS and lie within a neighborhood of the sun with a radius of 800 pc.

It was found [35] that the abundance of helium in the atmospheres of these B-stars increases along the MS, with the increase in He/H being greater for more massive stars and higher rotation velocities. At the start of the MS phase, these stars have an average helium abundance of *He/H* = 0.098 [37], and at the end of this phase He/H can be greater by more than a factor of two. In order to eliminate this evolutionary effect, we shall consider only the unevolved stars at the beginning of the MS phase. More precisely, we shall consider stars with relative ages $t/t_{MS} \leq 0.30$, where $t$ is the age of a star and $t_{MS}$ is the lifetime of a star with a given mass in the MS. In addition, we exclude stars with rapid rotation, limiting ourselves to stars with observed rotation velocities $v \sin i \leq 130$ km/s.

It was ultimately possible to select 18 early B-type MS stars that met these conditions from the list in Ref. 35. A plot of He/H as a function of $T_{eff}$ for these stars shows that there is no obvious trend in the helium abundance with $T_{eff}$ of the sort found for C, N, and O (Figs. 1-3). Thus, it seems that there is no need to invoke overionization. It is interesting to note that Gies and Lambert [2] also found no trend for He I, as opposed to C II, N II, and O II (see Fig. 11 in Ref. 2). Why is the situation for C, N, and O different? As noted above, in the non-LTE calculations it is necessary to include two ionization processes: collisional and radiative (i.e., photoionization). Helium atoms are 3-4 times lighter than C, N, and O atoms; thus, we may assume that collisional ionization of He I will predominate over photoionization processes, as opposed to the case of C II, N II, and O II.

6. Conclusion

The goal of this study has been to draw attention to the following fact, which is of primary importance but has been ignored until recently: C II, N II, and O II ions, lines of which are usually used to determine the C, N and O abundances in the atmospheres of early B-stars and late O-stars, are photoionized by far UV radiation, but there appear to be large errors in the theoretical UV fluxes employed in non-LTE calculations of the C II, N II, and O II lines.

In fact, an analysis of observations of early B-stars in this region shows that the theoretical UV fluxes calculated on the basis of standard models for stellar atmospheres can be significantly understated. This leads to an underestimate in the degree of ionization of C II, N II, and O II in the calculations and, ultimately, to abundances of C, N, and O that are too low. It is necessary to introduce additional ionization (overionization) in order to eliminate the systematic reduction in these abundances in the hottest B-stars, as well as in O-stars.

Until recently, standard models of stellar atmospheres, i.e., plane-parallel LTE models with hydrodynamic equilibrium, have been used for studying the C, N and O abundances in early B-stars and late O-stars. They have been used, first of all, for determining the fundamental parameters $T_{eff}$ and log *g* and, secondly, in non-LTE calculations of the lines of C II, N II, and O II. It turned out that the resulting C, N, and O abundances depend on the approach used to estimate $T_{eff}$ and log *g*.

On one hand, if $T_{eff}$ and log *g* for hot stars are found using the traditional method in which the effective temperature $T_{eff}$ is estimated on the basis of photometric indices and the acceleration of gravity in the atmosphere, *log g*, is determined from Balmer lines (or from parallaxes [7]), the resulting abundances of C, N, and O manifest a trend with $T_{eff}$. A systematic reduction in the abundance of carbon becomes noticeable for $T_{eff}$ > 18500 K and in the abundances of nitrogen and oxygen, for $T_{eff}$ > 26000 K. For the comparatively cool B-stars with lower $T_{eff}$, the C, N and O abundances are the same as the solar abundances on the average, but for the hotter stars with these $T_{eff}$, there is a deficit of about 0.2 dex in C, N, and O. In order to avoid trending with $T_{eff}$ it is necessary to introduce overionization of C II, N II, and O II for the comparatively hot stars, which leads to a reduction in the populations of the upper levels for the observed lines. A crude estimate shows that the populations of these levels should be lowered by roughly a factor of 1.6.

On the other hand, if $T_{eff}$ and log $g$ are found using an ionization balance technique where, for example, the abundance of C based on C II lines is fit to the abundance of C based on C III lines, then $T_{eff}$ and log $g$ are enhanced for the relatively hot stars with $T_{eff} \geq 24000$ K. Then the overestimate in $T_{eff}$ compared to the traditional method can be as much as 2300 K, and that in log $g$, as much as 0.5 dex. Both of these causes (an increase in both $T_{eff}$ and log $g$) automatically lead to enhanced ionization of C II, N II, and O II.

Therefore, when standard models for stellar atmospheres are used there is an ambiguity in the determination of $T_{eff}$ and log $g$ for hot stars and, thereby, an ambiguity in the C, N and O abundances obtained with them. Values of $T_{eff}$ and log $g$ found from photometric indices and Balmer lines cannot simultaneously satisfy the condition of ionization balance. The major reason for these disagreements is probably that the standard models give a reduced UV flux for hot stars that leads to a reduced degree of ionization of C II, N II, and O II in the calculations.

In order to solve this problem, it is necessary to shift to the use of more complicated but more realistic models for stellar atmospheres in studies of the chemical composition of early B- and O-stars. Some early success can be attained by going from plane-parallel to spherical models. Including the stellar winds and magnetic fields observed in some O- and B-stars in the calculations may also have some effect.